\documentclass[]{spie}  

\newcommand{\pseudodot}{{\lower 2.4pt\hbox{$\cdot$}}}

\usepackage{amsmath}
\usepackage{graphicx}
\usepackage[caption=false]{subfig}
\usepackage{float}
\usepackage{multirow,array,booktabs}

\DeclareUnicodeCharacter{2212}{-}
\newcommand{\ucsdp}{Department of Physics, University of California, San Diego, La Jolla, CA 92093}
\newcommand{\ucsda}{Department of Astronomy and Astrophysics, University of California, San Diego, La Jolla, CA 92093}
\usepackage{amsmath,amsfonts,amssymb}
\usepackage{graphicx}
\usepackage{threeparttable}

\usepackage[colorlinks=true, allcolors=blue]{hyperref}

\title{GPI 2.0: Exploring The Impact of Different Readout Modes on the Wavefront Sensor's EMCCD}

\author[a]{Clarissa R. Do Ó}
\author[b]{Saavidra Perera}
\author[b]{Jérôme Maire}
\author[b]{Jayke S. Nguyen}
\author[k]{Vincent Chambouleyron}
\author[b]{Quinn M. Konopacky}
\author[c]{Jeffrey Chilcote}
\author[d]{Joeleff Fitzsimmons}
\author[c]{Randall Hamper}
\author[d]{Dan Kerley}
\author[k]{Bruce Macintosh}
\author[d]{Christian Marois}
\author[f]{Fredrik Rantakyrö}
\author[g]{Dmitry Savranksy}
\author[d]{Jean-Pierre Veran}
\author[h]{Guido Agapito}
\author[i]{S. Mark Ammons}
\author[h]{Marco Bonaglia}
\author[j]{Marc-Andre Boucher}
\author[d]{Jennifer Dunn}
\author[h]{Simone Esposito}
\author[j]{Guillaume Filion}
\author[j]{Jean Thomas Landry} 
\author[d]{Olivier Lardiere}
\author[g]{Duan Li}
\author[e]{Alex Madurowicz}
\author[c]{Dillon Peng}
\author[i]{Lisa Poyneer}
\author[c]{Eckhart Spalding}

\affil[a]{\ucsdp}
\affil[b]{\ucsda}
\affil[c]{Department of Physics, University of Notre Dame, 225 Nieuwland Science Hall, Notre Dame, IN 46556, USA}
\affil[d]{National Research Council of Canada Herzberg, 5071 West Saanich Rd, Victoria, BC, V9E 2E7, Canada}
\affil[e]{Kavli Institute for Particle Astrophysics and Cosmology, Stanford University, Stanford, CA 94305, USA}
\affil[f]{Gemini Observatory, 670 N. A’ohoku Place, Hilo, HI 96720, USA}
\affil[g]{Sibley School of Mechanical and Aerospace Engineering, Cornell University, Ithaca, NY 14853, USA}
\affil[h]{Arcetri, Largo Enrico Fermi 5, I - 50125 Florence, Italy}
\affil[i]{Lawrence Livermore National Laboratory, Livermore, CA 94551, USA}
\affil[j]{Opto-Mécanique de Précision, 146 Bigaouette St. Quebec City, QC, Canada, G1K 4L2}
\affil[k]{Center for Adaptive Optics, University of California Santa Cruz, Santa Cruz, CA 95064, USA}

\authorinfo{Further author information: (Send correspondence to C.D.O) \\ E-mail:cdoo@ucsd.edu}

\authorinfo{Further author information: (Send correspondence to Clarissa R. Do Ó)\\Clarissa Do Ó: E-mail: cdoo@ucsd.edu}

\begin{document} 
\maketitle

\begin{abstract}
The Gemini Planet Imager (GPI) is a high contrast imaging instrument that aims to detect and characterize extrasolar planets. GPI is being upgraded to GPI 2.0, with several subsystems receiving a re-design to improve its contrast. To enable observations on fainter targets and increase performance on brighter ones, one of the upgrades is to the adaptive optics system.  The current Shack-Hartmann wavefront sensor (WFS) is being replaced by a pyramid WFS with an low-noise electron multiplying CCD (EMCCD). EMCCDs are detectors capable of counting single photon events at high speed and high sensitivity. In this work, we characterize the performance of the HNü 240 EMCCD from Nüvü Cameras, which was custom-built for GPI 2.0. Through our performance evaluation we found that the operating mode of the camera had to be changed from inverted-mode (IMO) to non-inverted mode (NIMO) in order to improve charge diffusion features found in the detector's images. Here, we characterize the EMCCD's noise contributors (readout noise, clock-induced charges, dark current) and linearity tests (EM gain, exposure time) before and after the switch to NIMO.
\end{abstract}

\section{INTRODUCTION}
Finding new exoplanets requires the development of state-of-the-art high contrast imaging instruments.
Directly imaged planets are much fainter than their host stars, which is why most high contrast imaging instruments require a coronagraph, a device that blocks the light from the star such that an off-axis signal, like the planet's light, can be detected \cite{Milli2017}. However, proper imaging of an off-axis source requires the coronagraph to be well-aligned with the starlight. Low-order wavefront aberrations, such as tip or tilt, can severely suppress the contrast capabilities of an instrument by allowing the starlight to ``leak" into the rest of the image.  This necessitates efficient wavefront measurement and correction using a highly sensitive wavefront sensor and deformable mirrors. \par
The Gemini Planet Imager (GPI) is a high contrast imaging instrument that operated at Gemini South for 6 years with the main goal of studying gas giant planet occurrence at wide orbits. Its goal was also to constrain whether these planets were in agreement with the formation via core-accretion or gravitational instability. Planets that form via core-accretion have lower entropy in their formation due to the formation of an accretion disk around the solid core of the forming planet \cite{Spiegel2012}. This lower entropy causes these planets to be fainter than their gravitational instability counterparts, where the unstable region in the disk collapses directly to form a planet that retains much of its initial entropy \cite{Marley2007}. Therefore the contrast requirements for finding both types of planets are quite demanding.\par

\begin{figure} [ht]
   \begin{center}
   \begin{tabular}{c} 
   \includegraphics[height = 7cm]{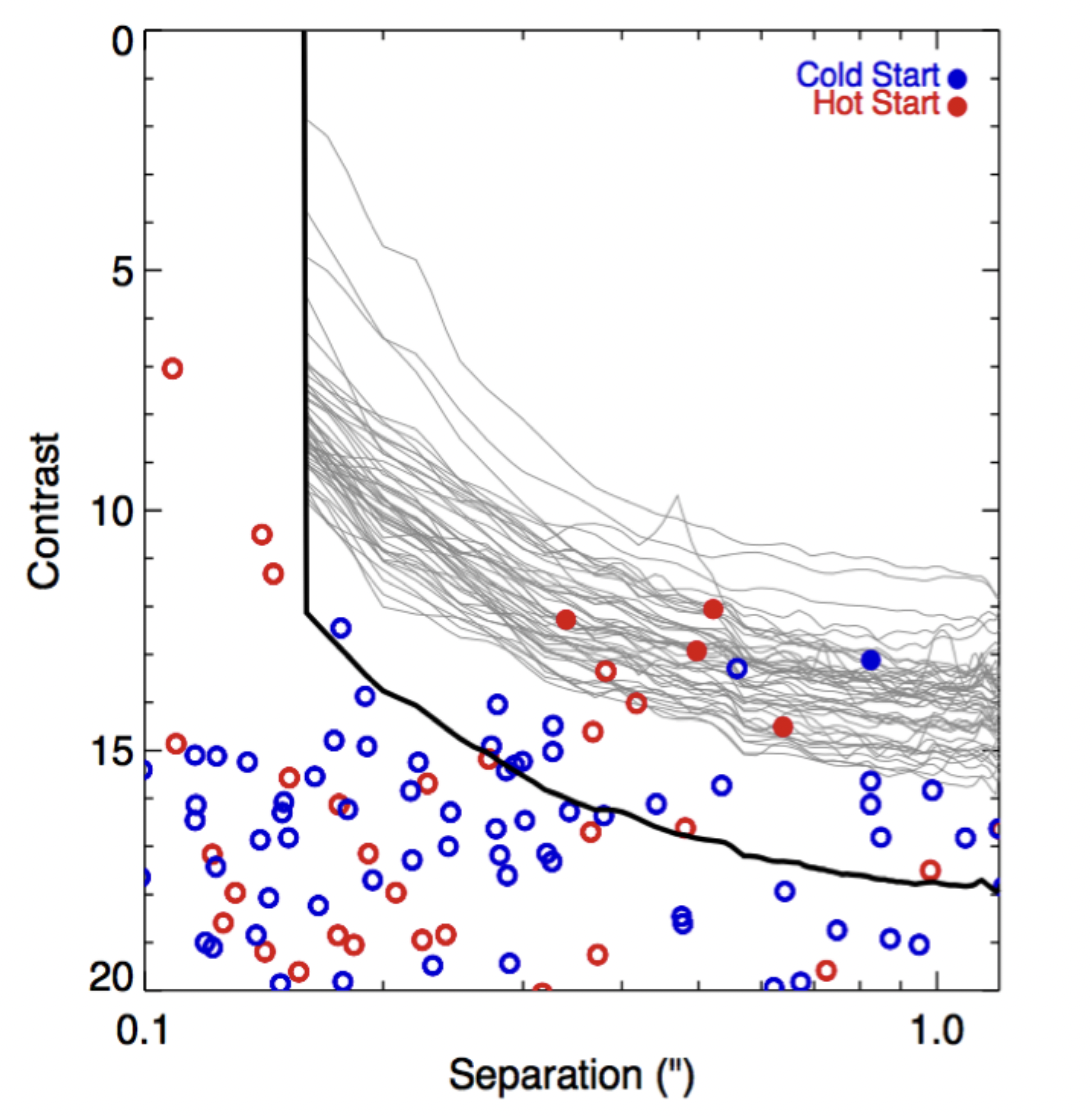}
   \end{tabular}
   \end{center}
   \caption[example] 
   {\label{fig:chilcote.png} 
GPI's contrast curve (in magnitudes) as a function of separation in arcseconds from the host star. Measured GPI 1.0 contrast curves are represented in light gray, with the GPI 2.0 predicted contrast curve represented in black. The GPI 2.0 will reach planets 3.2 magnitudes fainter than GPI 1.0, allowing for many more ``cold-start" and closer-in planets to be detected. The filled in circles are planets that were found by GPI 1.0's current set-up, while hollow circles represent planets that would fall below the current contrast curve for the host star. The planets come from a simulated exoplanet population. Figure from \cite{Chilcote2018}.}
   \end{figure}

The improvement in high contrast imaging technologies now allows for better contrasts at smaller separations from the host star, enabling not only the detection of closer-in planets but also for planets more consistent with the core accretion model  \cite{Chilcote2018}, \cite{Nielsen_2019}. GPI is going through an upgrade to become GPI 2.0, which is designed to achieve higher contrast than GPI 1.0. In order to achieve this goal, several subsystems are receiving upgrades, including the calibration unit, the coronagraphic system, the integral field spectrograph (IFS) and the adaptive optics system (AO). The predicted contrast after the upgrade to GPI 2.0 is shown in Figure \ref{fig:chilcote.png}. The figure shows that GPI 2.0 will unlock the detection of planets consistent with the cold-start models, as GPI 1.0 was already quite sensitive to hot-start planets \cite{Chilcote2018}, enabling detections at the peak of the expected giant planet population distribution \cite{Nielsen_2019}. \par

\subsection{The Gemini Planet Imager 2.0: Wavefront Sensor Upgrade} \label{pwfs}

\begin{figure} [ht]
   \begin{center}
   \begin{tabular}{c} 
   \includegraphics[width = 12cm]{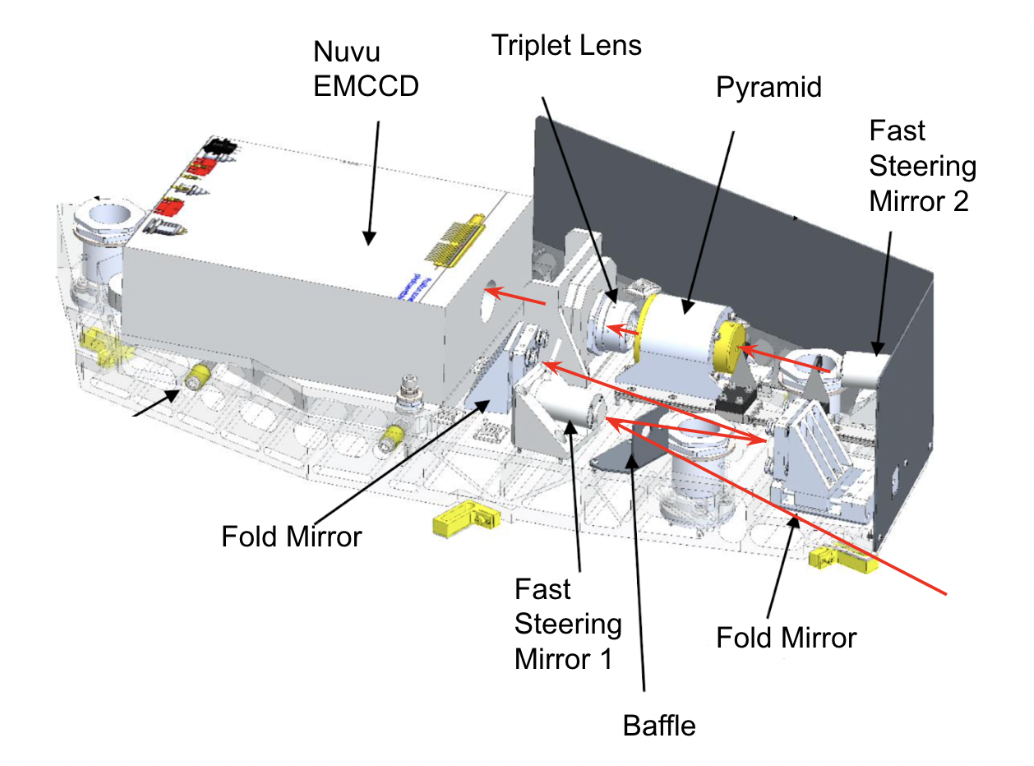}
   \end{tabular}
   \end{center}
   \caption[example] 
   {\label{fig:pwfs} 
GPI 2.0's pyramid WFS was designed by the Herzberg Astronomy and Astrophysics Research Center (HAA) and simulated by Stanford University, with the assembly taking place at UCSD. The design includes two fast steering mirrors, one for modulation (FSM 1) and one for tip/tilt adjustments (FSM 2). Between the two FSMs, there are two fold mirrors; the first fold mirror also acts as a focus stage. After FSM 2, the light goes through the double four-sided pyramid and through a camera lens before reaching the Nüvü EMCCD.}
   \end{figure}

AO subsystem upgrades occurred at the University of California, San Diego (UCSD) \cite{Perera2022}. GPI 2.0's pyramid wavefront sensor has higher sensitivity to low-order aberrations compared to its previous iteration (Shack-Hartmann) that will allow for the detection of fainter targets. 
GPI 2.0's wavefront sensor design is shown in Figure \ref{fig:pwfs}. The modulation around the tip of the pyramid occurs on the first fast steering mirror (FSM 1) in order to increase the dynamic range of the WFS. It also has a fold mirror which allows for focusing (focus stage) of the beam and a second fast steering mirror for tip and tilt adjustments. The light goes through the double four-sided pyramid and a triplet lens before hitting the detector, the Nüvü EMCCD.

\subsection{The EMCCD}

\begin{figure}
  \begin{center}
\centerline{\includegraphics[width=14cm]{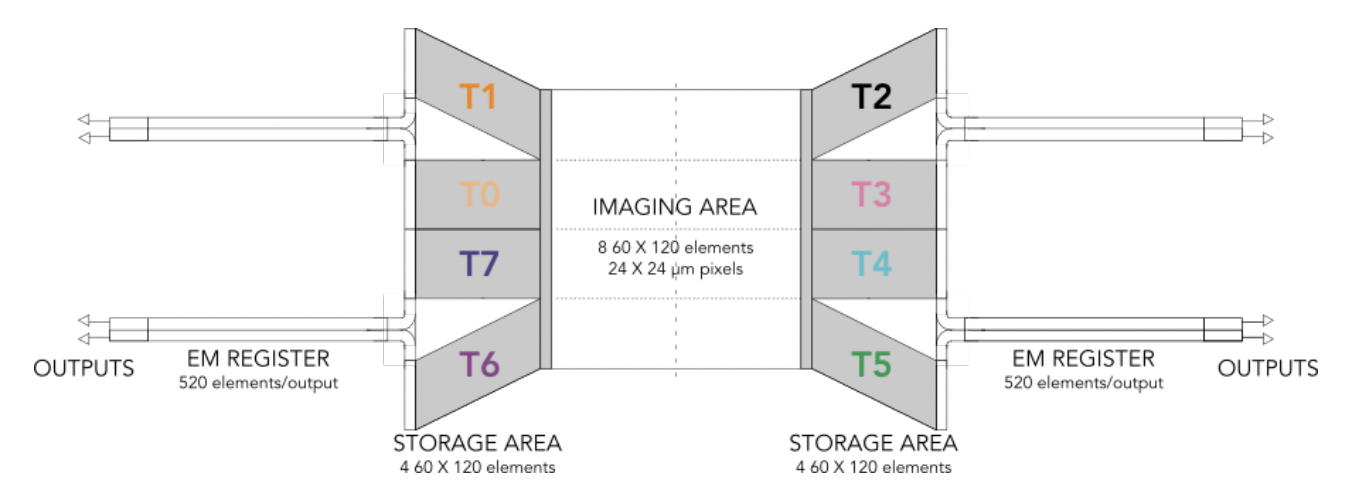}}
\caption{The EMCCD's chip. The imaging area is composed of 8 60x120 imaging areas ("outputs"), forming a 240x240 pixel image, storage areas and multiplication ("EM") registers. The camera has a nominal temperature operation of -45 $^{\circ}$C. Figure is from \cite{10.1117/12.2626240}}. 
\label{fig:emccdchip}
  \end{center}
\end{figure}

The GPI 2.0 wavefront sensor's detector is a NuVu Cameras electron-multiplying CCD (EMCCD) \cite{NuVuCameras}. EMCCDs are detectors capable of counting single photon events at high speed and high sensitivity. The EMCCD chip configuration is presented in Figure \ref{fig:emccdchip}. The detector has 8 different outputs. Much like a traditional CCD, the EMCCD turns photons into electrons via the photoelectric effect. However, unlike the traditional CCD, the EMCCD has an extra register called the multiplication register. Once the photons hit the silicon body of the chip, the electrons in the imaging area travel row by row to the storage area. This mechanism allows for the next frame to be taken while the previous one is being processed \cite{NuVuCameras}. Once in the storage section, the electrons travel to the multiplication register where hundreds of electrodes accelerate them, causing a phenomenon called impact ionization. Using high voltages, the captured electrons collide with the multiplication registers' silicon atoms, ripping an electron from the atom. This new electron then becomes part of the measured signal \cite{NuVuCameras}. \par
The EM Gain sets how much an electron signal will be multiplied by, which is achieved by changing the voltage in the multiplication register. This specific EMCCD chip, CCD220, has 8 outputs for a faster readout of these electrons, with 2 outputs sharing one multiplication register. EMCCDs are particularly useful for AO wavefront sensor systems because of their high sensitivity at a high operating speed. This EMCCD's operating speed of 3,000 frames per second (FPS) and high sensitivity combined with low noise allows it to keep up with the changing atmosphere and therefore for better corrections of the wavefront to high order aberrations \cite{10.1117/12.2626240}, \cite{doi:10.1142/S2251171719500016}. Our camera has a nominal temperature operation of -45 $^{\circ}$C.

\subsection{Inverted Mode, Non-Inverted Mode and Charge Diffusion}

After extensive testing at UCSD the EMCCD was found to have charge diffusion due to its ``inverted mode" set-up.
EMCCDs have two readout modes for operation: inverted vs. non-inverted mode (IMO and NIMO). The difference in the modes can be set by the voltages across the sensor. The IMO is at a lower voltage, generating ``holes" in the sensor's substrate, which recombine with dark current electrons before reaching the readout register \cite{NuVuCameras}. These holes, however, generate more clock-induced charges (CIC) during the vertical transfer of electrons to the readout register. This causes a significant decrease in the detector's dark current, but an increase in CIC. IMO was the previous mode of operation for this EMCCD. The iteration of tests in this mode is presented in an AO4ELT7 proceeding \cite{DoO2023}.
\par
IMO has previously been found to cause charge diffusion due to its reduction in the potential barrier between adjacent pixels \cite{Downing2015}. This reduction in the potential barrier between neighboring pixels caused a significant charge diffusion effect (blurring) in our images. The result also appeared to be wavelength-dependent, with shorter wavelengths showing slightly larger FWHMs in their PSFs. \par
The EMCCD was sent back to Nüvü cameras so its readout mode could be changed to NIMO, which would increase the potential barrier between adjacent pixels and thus improve the resolution of the detector. This mode is expected to decrease CIC events. However, it would also significantly increase dark current, although such an effect should not be expected to significantly affect the pyramid wavefront sensor's performance since the frame rate of operations is quite high (2k FPS), and the camera is cooled to -45 C to mitigate dark current.
\par
This report presents the results of EMCCD performance after its change from IMO to NIMO. Here we also compare the results in both cases.

\section{Camera Resolution}

\begin{figure}[hb!]
    \centering
    \includegraphics[width=.7\linewidth]{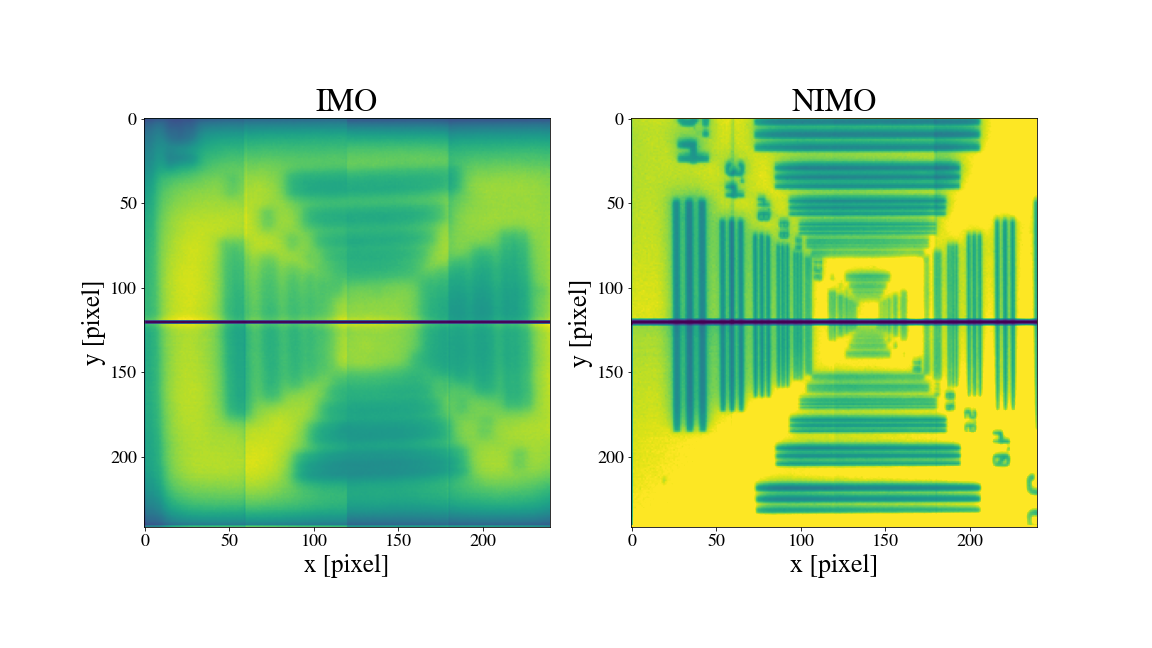}
    \caption{Qualitative comparison between the NIMO and IMO images of the resolution card. In both cases, a lens was used to focus the image on the camera. However, for IMO, the charge diffusion effect caused the image to look blurry.}
    \label{fig:nimovsimo}
\end{figure}

Due to the charge diffusion effect caused by IMO, we assess the change in image quality of the camera when the camera's operation mode was changed to NIMO. In order to do this, we image a standard resolution card by Thorlabs (NBS 1952) and analyze the line resolutions for our corresponding pixel size. \par
The camera possesses a 24 $\mu$m pixel size. The image of our resolution card is shown in Figure \ref{fig:nimovsimo} for the IMO and NIMO modes of operation. It is clear from the figure that changing the operation mode drastically changes the resolution capabilities of the camera.

\begin{figure}[hb!]
    \centering
    \includegraphics[width=.7\linewidth]{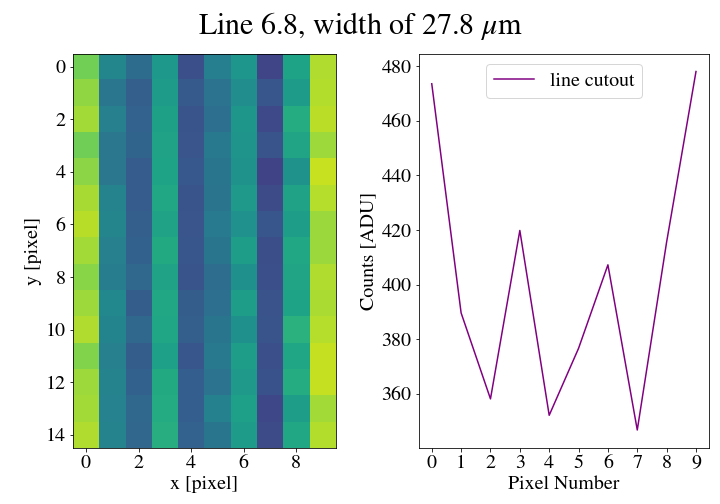}
    \caption{Cross cut of line 6.8, which has a linewidth of 73.5 $\mu$m. For our magnification set-up, it presents a width of 27.8 $\mu$m, slightly above one pixel width.}
    \label{fig:crosscut}
\end{figure}

In order to verify that the camera is operating within the expected resolution, we must verify that the pixel size resolution is achieved. In order to do this, we use the set lines on the resolution grids, which have specific sizes in $\mu$m. We focus on line 6.8 for our analysis, which has a linewidth of 73.5 $\mu$m. Given the magnification factor of , these lines should have a size of 27.8 $\mu$m, which is about 1.2 pixels. Our results of a cross cut for these lines are shown in Figure \ref{fig:crosscut}. We fit a curve to the image of the lines' cross cut and find that the actual width is closer to 1.3 pixels, which is slightly above but near the expected width given the camera's resolution.

\section{Conducted Tests}
\begin{figure}[h]
    \centering
    \includegraphics[width=.9\linewidth]{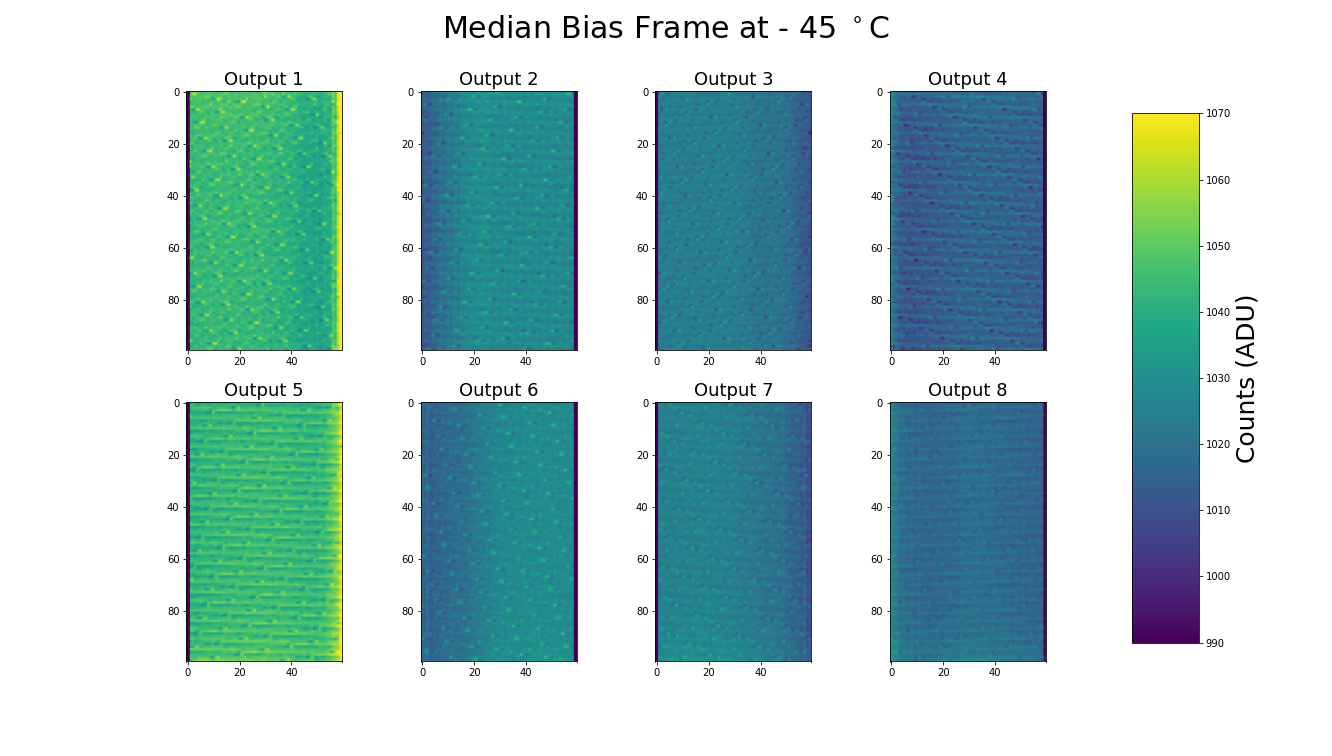}
    \caption{Representation of the 8 outputs of our camera. We number them from 1-8 as shown in the Figure. In order to better characterize the EMCCD, we always separate our results into the 8 outputs.}
    \label{fig:outputrep}
\end{figure}

For all of our conducted tests, we separate our results into each of the 8 outputs of our EMCCD, as was done in the previous iterations of tests \cite{DoO2023}.  Separating the results allows for a better characterization of the individual outputs of the EMCCD. We represent the 8 outputs for a median bias frame at -45 $^{\circ}$C in Figure \ref{fig:outputrep}.

\subsection{Readout Noise}
We repeat our readout noise procedure described in the previous camera test report, where the camera was operating in its inverted mode (IMO) \cite{DoO2023}, but now we compare the results to the NIMO operating mode. For each output, we subtract the median of dark frames from the 1,000 dark frames and obtain the standard deviation of the 1,000 frames. We then multiply the standard deviation frame by the K-Gain set for each exposure series and divide by the EM gain of 5,000 to obtain units of electrons. We then obtain the median of these values for each output. \par
We represent the median readout noise for each output in Table \ref{tbl:1}. We found that it the median readout noise does not significantly change from what it was when the camera was on IMO. Despite the requirement stating 0.1 e-, all of the contrast simulations were performed expecting 0.4 e- of readout noise. Therefore, all of the outputs are below the values used in simulations for the wavefront sensor's expected performance.

\begin{table}[ht]
\caption{Median Readout Noise of the EMCCD [e-] for each Detector Output} 
\label{tbl:1}
\begin{center}       
\begin{tabular}{|l|l|l|} 
\hline
\rule[-1ex]{0pt}{3.5ex} Output & IMO (Before) & NIMO (After)\\
\hline
\rule[-1ex]{0pt}{3.5ex} 1 &  0.169787 & 0.166559 \\
\hline
\rule[-1ex]{0pt}{3.5ex} 2 &  0.167197 & 0.168746 \\
\hline
\rule[-1ex]{0pt}{3.5ex} 3  &   0.139553  & 0.141459 \\
\hline
\rule[-1ex]{0pt}{3.5ex} 4  &  0.150889 & 0.145825  \\
\hline
\rule[-1ex]{0pt}{3.5ex} 5  &  0.132648 & 0.132513 \\
\hline 

\rule[-1ex]{0pt}{3.5ex} 6  &  0.103851 & 0.103713 \\
\hline 
\rule[-1ex]{0pt}{3.5ex} 7  &  0.071985  &  0.0746640\\
\hline 
\rule[-1ex]{0pt}{3.5ex} 8  & 0.127600 & 0.125917  \\
\hline 
\end{tabular}
\end{center}
\end{table}

\subsection{Dark Current}

We also test the dark current present in our detector and compare it to our previous results when the camera was in the inverted mode of operation. We test this at EM gain of 5,000. We measure the dark current by changing the exposure time in our detector and taking dark frames. The dark frames are all bias subtracted with matching EM gain and all of the frames are taken at -45 $^{\circ}$C. The exposure times used are 1, 2, 4, 8 and 10 seconds. We calculate the dark noise as follows: we first subtract bias frame from mean of each exposure time for each output, then take the mean for the pixels in each output, obtaining units of [ADU/pix/fr]. Finally we transform from ADU to e- using the EM and k-gain values.

\begin{figure}
    \centering
    \subfloat[\centering ]{{\includegraphics[width= 8cm]{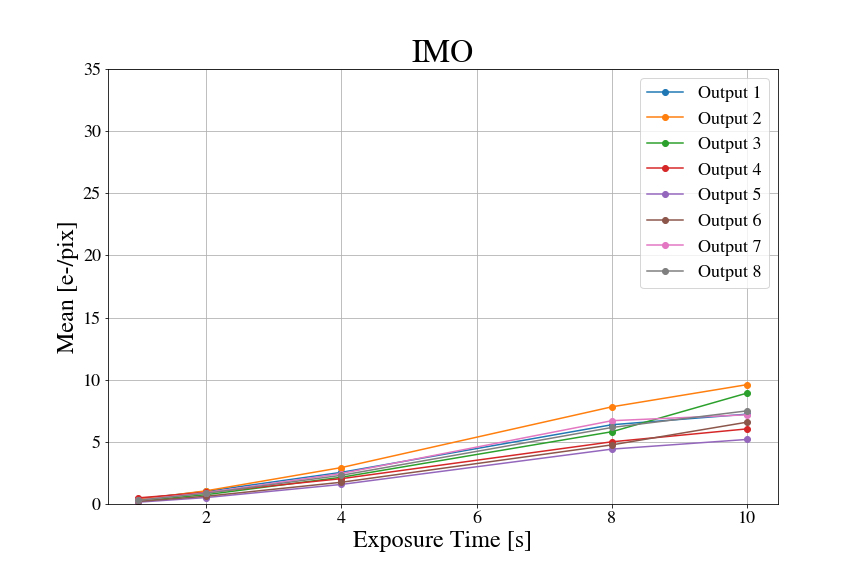} }}%
    \qquad
    \subfloat[\centering ]{{\includegraphics[width=8cm]{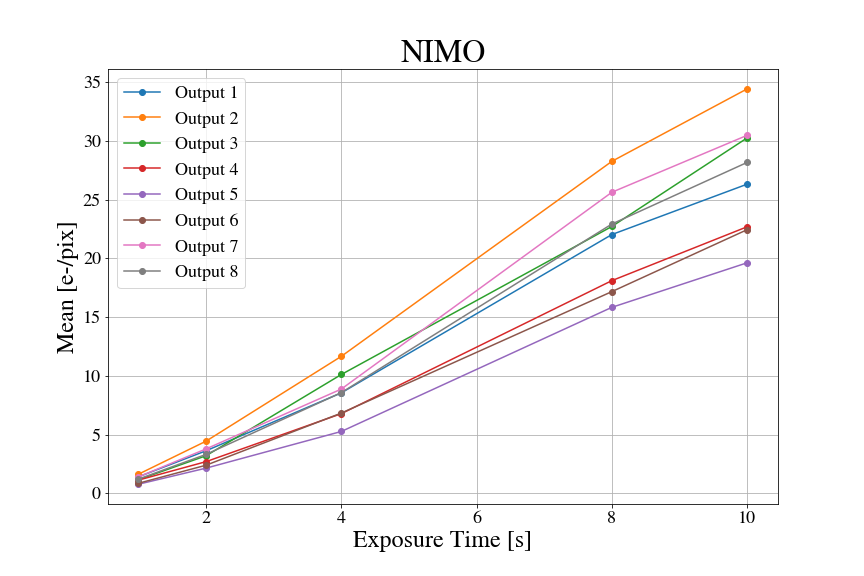} }}%
    \caption{The dark current in e-/pix for the mean frame of our EMCCD at EM gain of 5,000, for IMO (a) and NIMO (b) at -45$^{\circ}$C.}%
    \label{fig:darknoise}
\end{figure}

\begin{figure}[h]
    \centering
    \includegraphics[width=.8\linewidth]{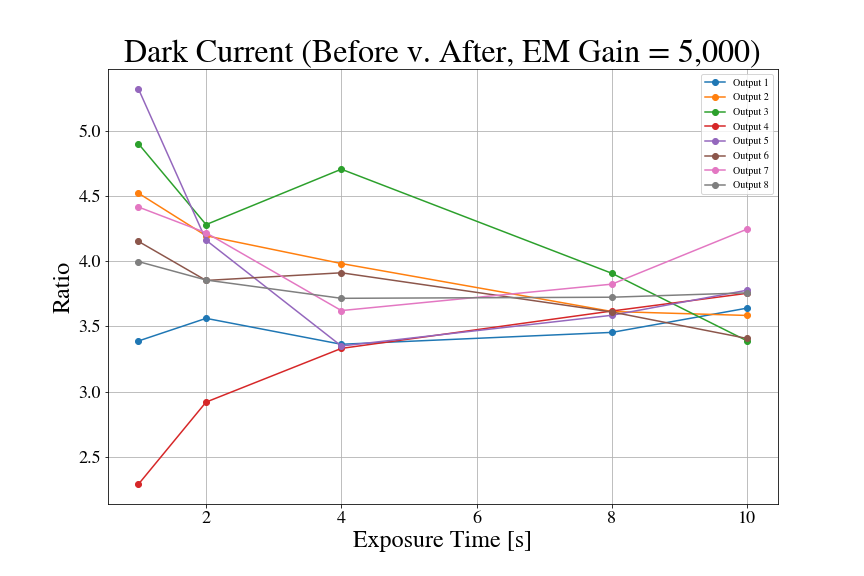}
    \caption{The dark current ratio for NIMO/IMO (After/Before). }
    \label{fig:darkcurrentratio}
\end{figure}

Our results are presented in Figure \ref{fig:darknoise}. We note that the dark current is about 3.8x (median) larger in NIMO than in IMO. A larger dark current in NIMO is expected. We plot the ratio for different exposure times in Figure \ref{fig:darkcurrentratio}.

\begin{figure}[h]
    \centering
    \includegraphics[width=\linewidth]{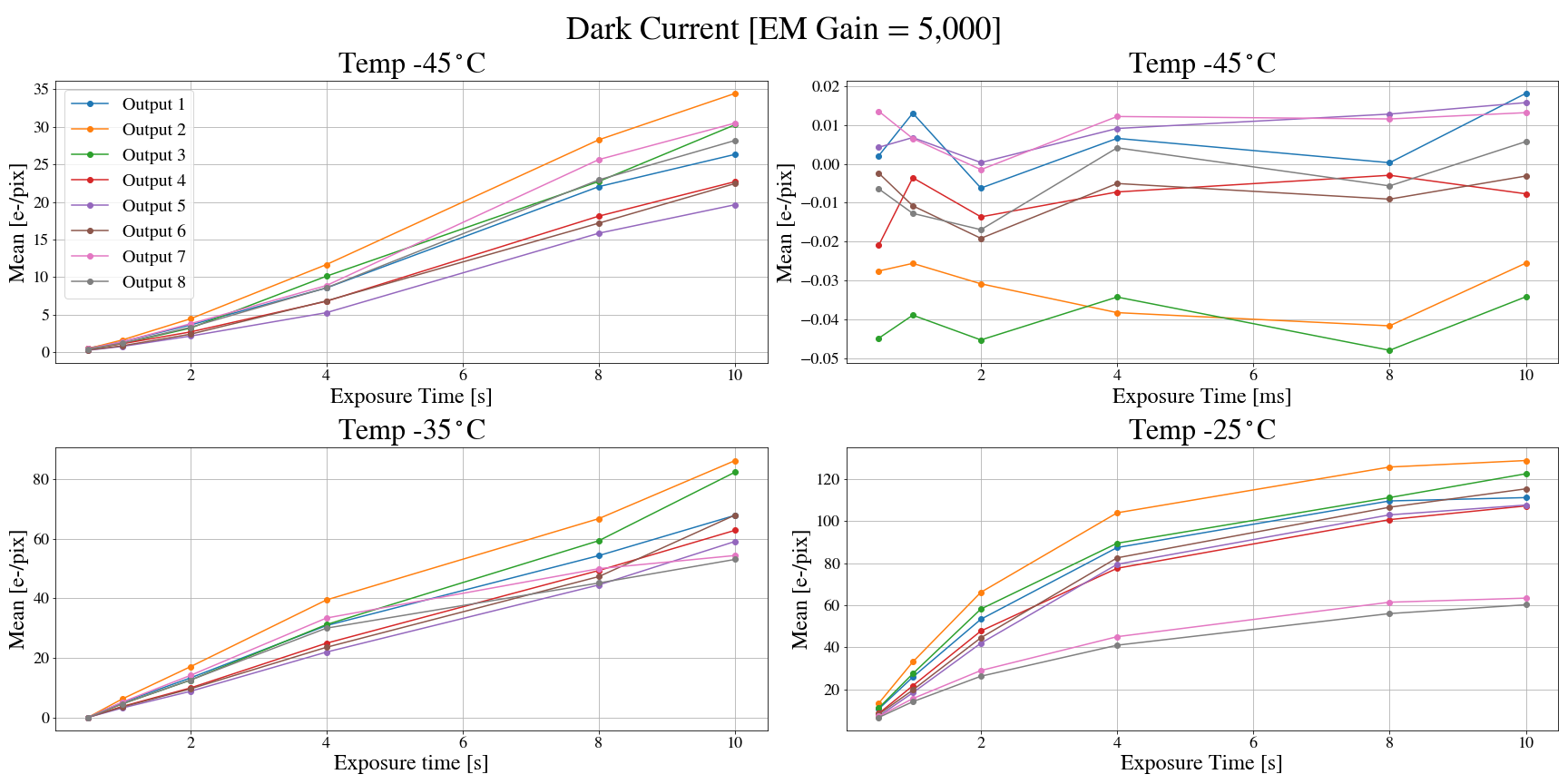}
    \caption{The dark current at different temperatures and exposure time regimes for EM Gain of 5,000. We note that for the top right panel, where the exposure times are low and close to our operation levels, the negative counts are due to the bias subtraction of the frames.}
    \label{fig:darkcurrenttemp}
\end{figure}

Finally, we test our dark current at exposure times for different temperatures (for -45 $^{\circ}$C, we test in the second and millisecond regime). We verify that in the ms regime, at -45 $^{\circ}$C, which is where we will be operating our wavefront sensor, dark current is low even in NIMO mode (see Figure \ref{fig:darkcurrenttemp}, top right panel). We note that the dark current is temperature dependent, as is expected, where we find saturating levels for high EM gain and exposure time for -25 C.  We show our results in Figure \ref{fig:darkcurrenttemp}.

\subsection{Clock-Induced Charges} \label{cic}

We also test the clock-induced charges (CIC) of the EMCCD. The CIC is a source of noise in the EMCCD where false counts are created when the photoelectrons travel in the EM register \cite{NuVuCameras}. We repeat the procedure performed in our previous tests \cite{DoO2023}. We set the EMCCD to ``photon counting mode", where we take dark exposures at the max frame rate of 0.33 ms, and bias subtract them. Then, we obtain frames where all of the pixels should have ``0" counts unless a CIC event occurs, in which case the pixel will have a ``1" count.\par
The CIC average for each output is calculated by summing the CIC events for every output over the number of frames and then dividing the sum by the number of frames and pixels in each output (120 x 60).
Our results are shown in comparison to IMO in Table \ref{tbl:2}.

\begin{table}[ht]
\caption{Median CIC for each Detector Output} 
\label{tbl:2}
\begin{center}       
\begin{tabular}{|l|l|l|} 
\hline
\rule[-1ex]{0pt}{3.5ex} Output & IMO (Before) & NIMO (After) \\
\hline
\rule[-1ex]{0pt}{3.5ex} 1 & 0.000695 & 0.001242 \\
\hline
\rule[-1ex]{0pt}{3.5ex} 2 &  0.001129 & 0.001863 \\
\hline
\rule[-1ex]{0pt}{3.5ex} 3  &   0.000396 & 0.000621\\
\hline
\rule[-1ex]{0pt}{3.5ex} 4  &   0.000536  & 0.000828\\
\hline
\rule[-1ex]{0pt}{3.5ex} 5  &  0.000204 & 0.000207 \\
\hline 
\rule[-1ex]{0pt}{3.5ex} 6  &  0.000313 &  0.000621\\
\hline 
\rule[-1ex]{0pt}{3.5ex} 7  &  0.000449 & 0.000621 \\
\hline 
\rule[-1ex]{0pt}{3.5ex} 8  & 0.000425 & 0.000621 \\
\hline 
\end{tabular}
\end{center}
\end{table}

\subsection{Flat Field Tests}
We conduct flat fielding tests for our EMCCD. For this, we utilize the Newport 819D-SL-3.3 integrating sphere, used in
conjunction with a Newport 6332 quartz tungsten halogen lamp, a white light lamp operated at 50 W and a Newport 60043 socket
adaptor. The bulb allows for the change in light intensity using a Kikusui Stabilized power supply (Model PAB 8-2.5). We test the EM gain, exposure time and light level linearities for the EMCCD. For all of our tests, we subtract the bias frame with matching EM gain and use the median of a cube with 300 frames. 

\subsubsection{EM Gain Linearity}

\begin{figure}[ht]
    \centering
    \includegraphics[width=\columnwidth]{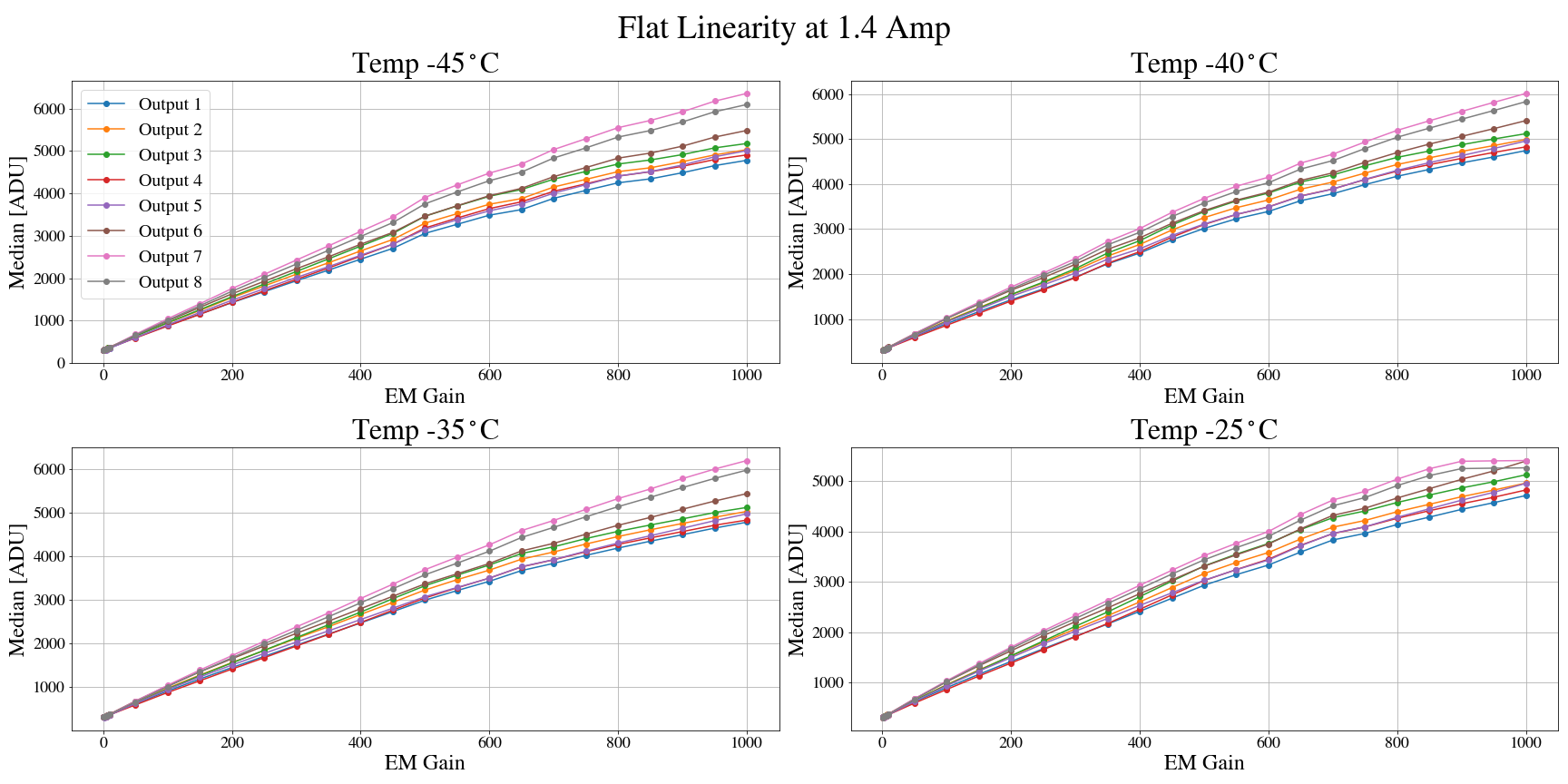}
    \caption{EM Gain Linearity test with matching light levels but varying temperatures for the EMCCD. The median of counts is shown as a function of EM gain. }
    \label{fig:gainhigh}
\end{figure}
We first test the linearity with changing EM gain for the camera at higher light levels (1.4 Amps). We test this at temperatures of -45, -40, -35 and -25 $^{\circ}$C up to 1000 EM gain. We do not go further than that such that we do not saturate all the outputs in the EMCCD, as that can damage the detector. We plan on operating the PWFS at -45 $^{\circ}$C, however this test allows us to characterize the camera's dependence on temperature. We subtract the bias with matching EM gain from each cube with 300 frames, then obtain the median of each output. We show our results in Figure \ref{fig:gainhigh}.

 We find that for Outputs 7 and 8 there is a saturation "cap" at higher temperatures (mainly -25 $^{\circ}$C). The same behavior was also present in IMO \cite{DoO2023}. \par 
 
\begin{figure}[ht]
    \centering
    \includegraphics[width=\columnwidth]{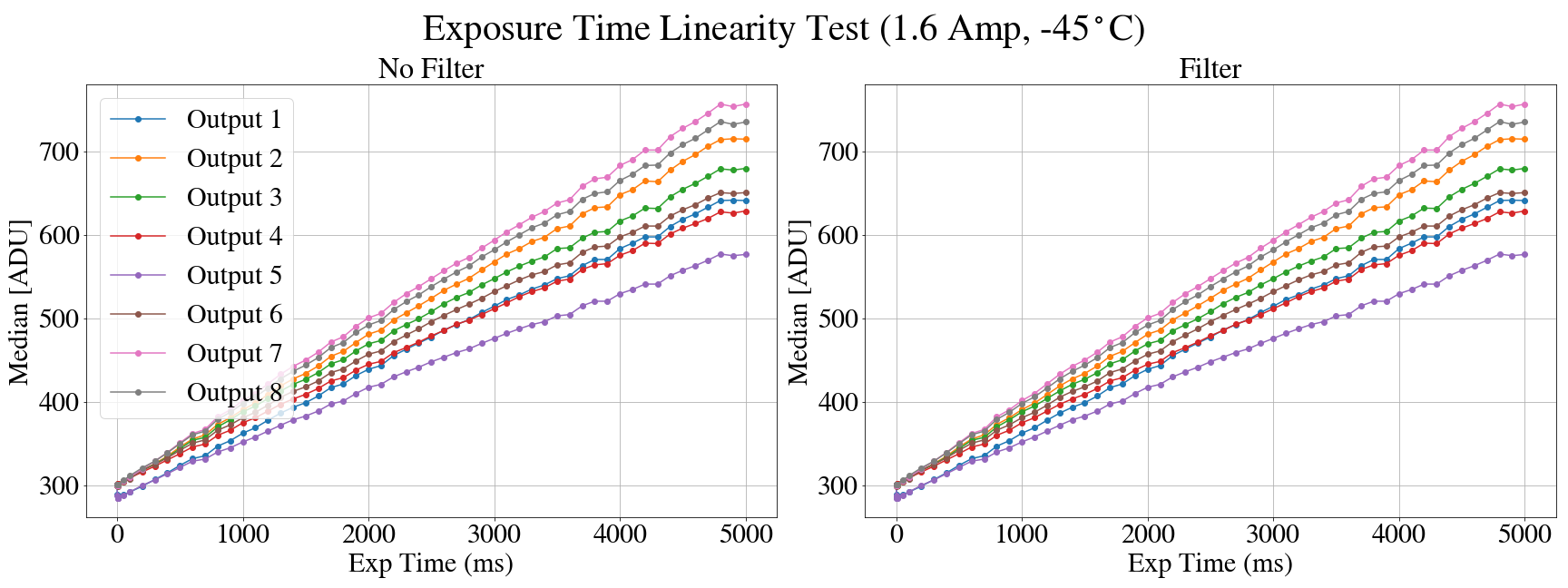}
    \caption{EM Gain Linearity test with lower light levels (1.1 and 1.2 Amp). The x-axis shows the EM gain values up to 5,000 while the y-axis shows the median of counts in ADU. }
    \label{fig:gainlow}
\end{figure}

 We then test the EM gain linearity at a lower light level, at 1.1 Amps and 1.2 Amps (up to 5,000 EM gain). We verify that at lower light levels we can obtain a linear behavior for the EM gain. We show our results in Figure \ref{fig:gainlow}.
 
\subsubsection{Exposure Time and Light Level Linearity}
\begin{figure}[ht!]
    \centering
    \includegraphics[width=\columnwidth]{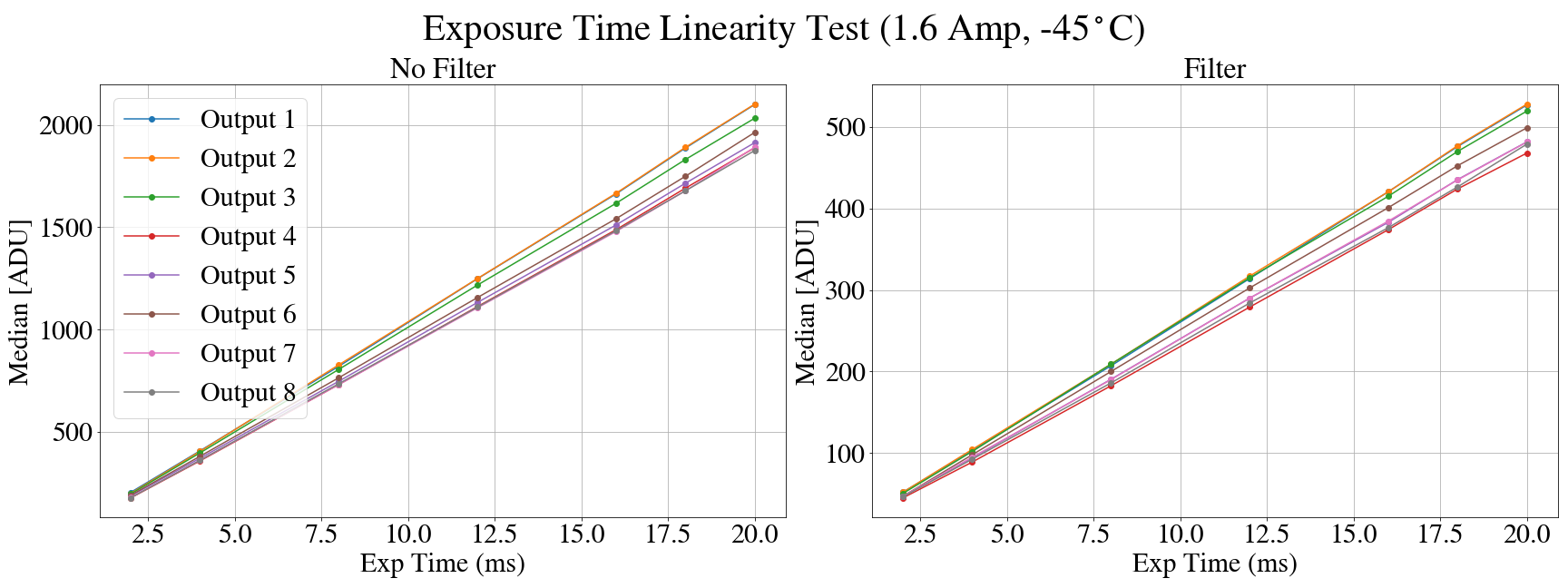}
    \caption{The median counts for flats with (left) and without (right) an ND filter as a function of exposure time in ms.}
    \label{fig:exptime}
\end{figure}

We test the exposure time linearity of our EMCCD. We perform this test for two light levels: by setting our flat lamp to 1.7 Amps and by keeping the level at 1.6 Amps but placing a Thorlabs Neutral Density (ND) 0.6 filter in front of the camera. This should decrease light levels by a factor of $\approx$ 4. Our exposure times used are 2, 4, 8, 12, 16, 18 and 20 ms. 
Results are presented in Figure \ref{fig:exptime}. We find that for both light levels the behavior all outputs is linear with changing exposure time.

\begin{figure}[ht!]
    \centering
    \includegraphics[width=.8\linewidth]{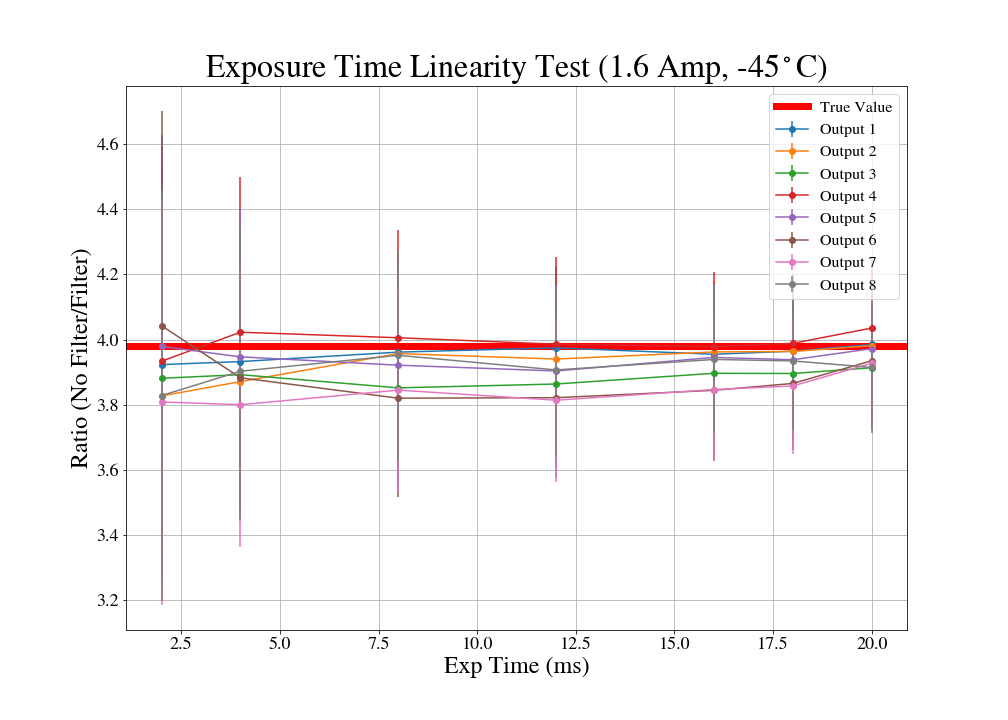}
    \caption{The ratio of light levels for flat fields with and without an ND filter. The true value is plotted as the red horizontal line, while measured levels are shown as scatter points with error bars for each output. Error bars are calculated using the Poisson noise for each output.}
    \label{fig:ratiolevel}
\end{figure}

Our results remain unchanged from the previous readout mode of the camera.
\subsubsection{Light Level Ratios}
We test that the counts obtained by the EMCCD correspond to the expected ratio given by our filter value (of 0.6, which corresponds to a decrease in light levels by a factor of 3.981). In order to do that, we plot the ratio of counts given by the two individual light levels at corresponding exposure times. Our results are presented in Figure \ref{fig:ratiolevel}.

Our results remain unchanged from the previous readout mode of the camera.

\subsection{Multiple Regions of Interest and Binning}

\begin{figure}[h]
    \centering
    \includegraphics[width=\linewidth]{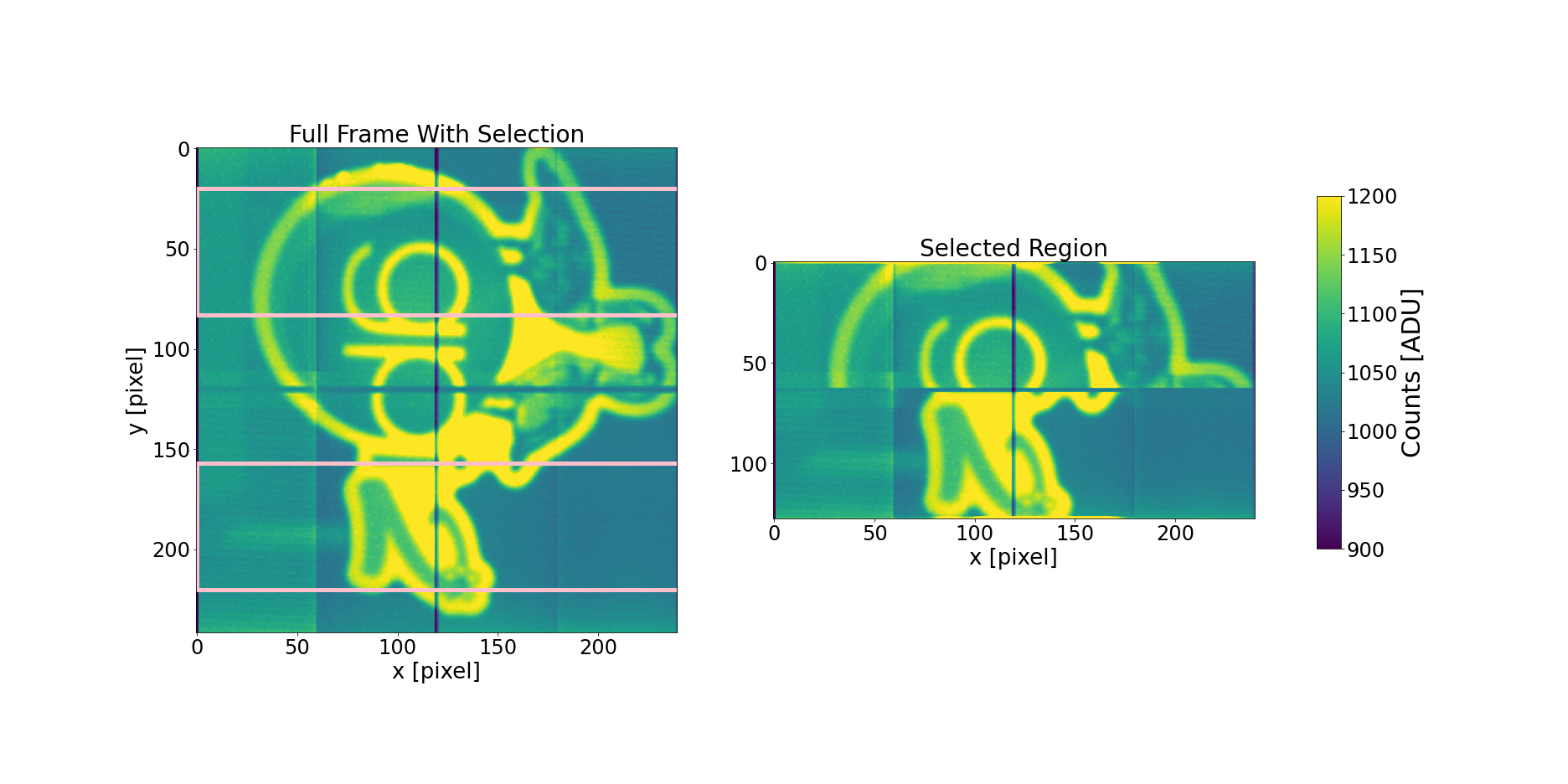}
    \caption{The multiple regions of interest of our EMCCD.}
    \label{fig:mroi}
\end{figure}

\begin{figure}[ht]
    \centering
    \includegraphics[width=0.9\linewidth]{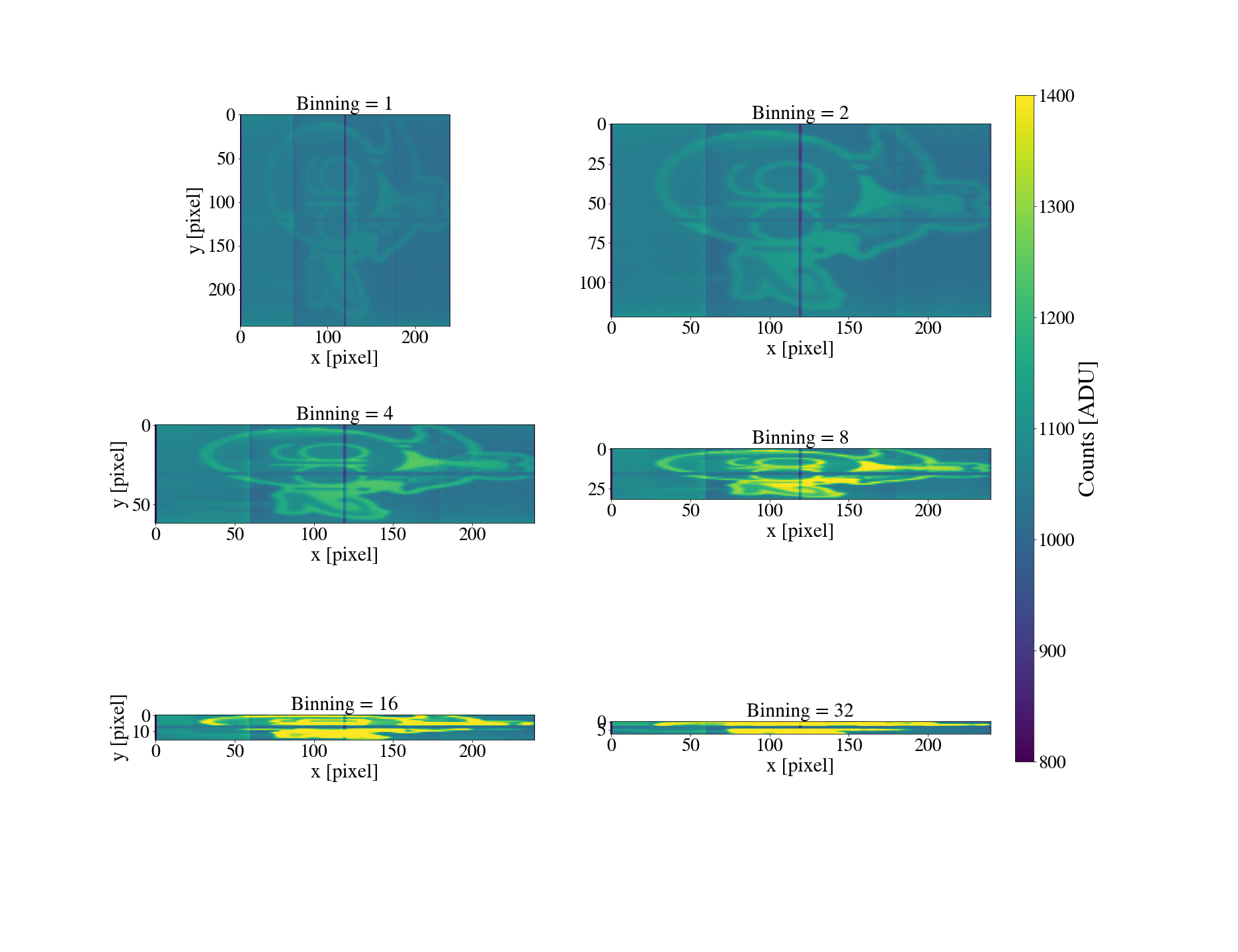}
    \caption{The binning feature of our EMCCD.}
    \label{fig:binning}
\end{figure}

We do not expect the multiple regions of interest (mROI) functionality and the binning of the EMCCD to be affected by the change in operating mode, but present the results for completeness. We repeat the test procedure from our previous iteration \cite{DoO2023}. The results are shown in Figure \ref{fig:mroi} and \ref{fig:binning}. We use an image of the GPI 2.0 logo for facilitating the visualization of the features. 

\section{CONCLUSION \& FUTURE WORK}

\begin{table}
\caption{Table with requirement tests for the EMCCD and their results.} 
   \begin{threeparttable}
   \begin{center}
\label{tbl:3}

\begin{tabular}{@{} p{4cm}|p{4cm}|p{4cm}|p{1cm} @{}}

 \hline
 Test & Requirement & Results & Pass or Fail? \\ [0.5ex] 
 \hline\hline
Readout Noise &  $<$ 0.1 e- readout noise at 3,000 FPS and EM gain of 5,000 & Median of 0.137 e- & Pass  \\
\hline
Dark Current &  $<$ 0.01 e-/pix/frame dark noise at 3,000 FPS & Median of 3.8x higher for longer exposures with NIMO, $<$ 0.008 e-/pix for all outputs at 2,000 FPS and EM Gain 5,000 & Pass  \\
\hline
Binning  & Binning Capability of 1, 2, 4, 8, 16, 32x & Binning Capability was achieved for all values & Pass \\
\hline
mROI  &  Configure the mROI for the 4 pupils of the EMCCD & mROI region was achieved with correct pixel offset and location & Pass  \\
\hline
Full Frame Rate & Full Frame Rate of 3 kHz & Full Frame Rate not achieved with SDK/GUI & $Fail^{*}$ \\
\hline
EM Gain Linearity  & EM gain functionality from 1 - 5,000 & Linearity Found for EM gain up to 5,000 when detector is not saturated & Pass \\
\hline
Exposure Time and Light Level Linearity  & The behavior of counts for exposure time and light levels must be linear & Linearity is achieved for all outputs & Pass \\
\\
 \hline

\end{tabular}
\begin{tablenotes}
   \item[*] Pending serial command test to achieve full frame rate.  
  \end{tablenotes}
\end{center}
   \end{threeparttable}
\end{table}
\label{sec:conclusion}
Our final performance table for the EMCCD is presented in Table \ref{tbl:3}. We find that the camera's performance in readout noise, mROI, CIC, binning, EM Gain linearity, exposure time linearity remain the same as before the change to NIMO readout mode. The dark current is about 3.8x higher than before, as expected for NIMO. However, in our mode of operations (low temperature and low exposure time), the difference in dark current is essentially negligible.  The image quality is improved and we are now achieving the expected resolution. This will allow us to obtain the required performance for the GPI 2.0 PyWFS.

\acknowledgments 
GPI 2.0 is funded by in part by the Heising-Simons Foundation through grant 2019-1582. The GPI project has been supported by Gemini Observatory, which is operated by AURA, Inc., under a cooperative agreement with the NSF on behalf of the Gemini partnership: the NSF (USA), the National Research Council (Canada), CONICYT (Chile), the Australian Research Council (Australia), MCTI (Brazil) and MINCYT (Argentina). Portions of this work were performed under the auspices of the U.S. Department of Energy by Lawrence Livermore National Laboratory under Contract DE-AC52-07NA27344.

\clearpage
\bibliography{report} 
\bibliographystyle{spiebib} 

\end{document}